\documentclass[conference]{IEEEtran}

\IEEEoverridecommandlockouts

\usepackage{todonotes}
\usepackage{listings}
\usepackage{amsmath,amsfonts}
\usepackage{soul}
\usepackage{caption}
\usepackage{multirow}
\usepackage{graphicx}
\usepackage{xcolor}
\usepackage{booktabs}
\usepackage{hyperref}
\usepackage{xurl}

\let\realurl\url
\renewcommand{\url}[1]{%
	\realurl{#1}%
	\wlog{URLX #1 }%
}

\begin{document}

%\title{Flaky Test Sanitisation via Assumption Inference On-The-Fly  with Application to Testing with Network Dependencies}
\title{Flaky Test Sanitisation via  On-the-Fly Assumption Inference for Tests with Network Dependencies}

%\maketitle

\author{
	\IEEEauthorblockN{Jens Dietrich\IEEEauthorrefmark{1}, Shawn Rasheed\IEEEauthorrefmark{2}, Amjed Tahir\IEEEauthorrefmark{2}}
	%\vspace{0.05in}
	\IEEEauthorblockA{\IEEEauthorrefmark{1}Victoria University of Wellington, Wellington, New Zealand}
	\IEEEauthorblockA{\IEEEauthorrefmark{2}Massey University, Palmerston North, New Zealand \\  jens.dietrich@vuw.ac.nz; s.rasheed@massey.ac.nz; a.tahir@massey.ac.nz}
	% \IEEEauthorblockA{\IEEEauthorrefmark{3}Massey University. Palmerston North, New Zealand. \emph{S.Rasheed@massey.ac.nz}}
}
\maketitle

\begin{abstract}

Flaky tests cause significant problems as they can interrupt automated build processes that rely on all tests succeeding and undermine the trustworthiness of tests. Numerous causes of test  flakiness have been identified, and program analyses exist to detect such tests. Typically, these methods produce advice to developers on how to refactor tests in order to make test outcomes  deterministic.

We argue that one source of flakiness is the lack of assumptions that precisely describe under which circumstances a test is meaningful. We devise a sanitisation technique that can isolate flaky tests quickly by inferring such assumptions on-the-fly, allowing automated builds to proceed as flaky tests are ignored. We demonstrate this approach for Java and Groovy programs by implementing  it as extensions for three popular testing frameworks (JUnit4, JUnit5 and Spock) that can  transparently inject the inferred assumptions. If JUnit5 is used, those extensions can be deployed without refactoring project source code.

We demonstrate and evaluate the utility of our approach using a set of six popular real-world programs, addressing known test flakiness issues in these programs caused by dependencies of tests on %networked resources
network availability. We find that our method effectively sanitises failures induced by network connectivity problems with high precision and recall.

%We conceptually underpin our work by distinguishing between two notions of flakiness, the traditional notion of  \textit{flakiness} that is based on the variability of test results, and the new proposed notion of \textit{strong flakiness} that focusses on the impact relevant changes in test results have on the results of automated builds.

\end{abstract}

%\keywords{flaky tests, non-determinism, regression testing}


%\todo[inline]{Shawn: Circumstances/factors that are outside the control of the test/application. \\Jens: while this makes sense for the network / environmental scenario, a developer could control test dependencies.}

\section{Introduction}

Flaky tests, i.e., tests with non-deterministic outcomes, are a widely known phenomena that causes significant problems in industry~\cite{googleFlaky2016,harman2018start,raine2020reducing,Agarwal2021Handling}. Modern software development processes heavily depend on a high level of automation, and for many projects,  automated regression testing remains the main means to demonstrate fitness for purpose. Therefore, it is desirable and often necessary for automated continuous integration (CI)  that all tests succeed. With flakiness, this process becomes brittle -- even if all tests succeed, it is unclear whether this is something that can be trusted, or whether tests have only accidentally succeeded.

%A test can only provide useful feedback if it has the same outcome (pass or fail) on every execution for the same version of code. Tests with non-deterministic outcomes (\textit{``flaky tests``}) are tests that may pass in some runs and fail on other runs. Test flakiness is a widely known problem that causes significant problems in industry~\cite{googleFlaky2016,harman2018start,raine2020reducing,Agarwal2021Handling}. Modern software development processes heavily depend on a high level of automation, and for many projects,  automated regression testing is the main mean to demonstrate correctness. Therefore it is desirable for automated CI/CD processes that tests succeed. With flakiness, this approach becomes brittle -- even if all tests succeed, it is unclear whether this is something that can be trusted, or whether tests succeed by chance.

%Failing automated tests usually lead to build failure, and they are the major cause for failing continuous integration (CI) pipelines~\cite{beller2017oops,durieux2020empirical}. There are, however, exceptions where some tests are permitted to fail~\cite{rogers2004scaling,van2007sisyphus,staahl2014modeling}.  
Dedicating resources to fix flakiness is  challenging as it is not clear to decision makers whether flakiness is caused by a fault in the program under test,  the test itself, or the test environment or configuration that is being used.  Fowler therefore suggested a two-staged process to deal with flaky tests~\cite{fowler2011eradicating}:

  \begin{quote}
	\textit{``Place any non-deterministic test in a quarantined area. (But fix quarantined tests quickly.)''}
\end{quote}

This reflects the situation developers face when they have to deal with flaky tests. For instance, Raine reports about the practices followed at GitHub~\cite{raine2020reducing}:

\begin{quote}
	\textit{``When we set out to build this new system, our intent wasn't to fix every flaky test or to stop developers from introducing new flaky tests. Such goals, if not impossible, seemed impractical. ... Rather, we set out to manage the inevitability of flaky tests.''}
\end{quote}

% This reflects the situation developers face when they have to deal with flaky tests, as expressed in this post~\cite{sudarshan2012nomore}:
%\begin{quote}
%	\textit{Lets say you have identified a flaky test. In fact, you might have identified, say, 23 of them. You cannot obviously fix all of them. That would take a lot of time.
%	In our case, we didn’t want to delete them, but at the same time, we did not want to infect our build with the disease of redness. A good middle ground was to Quarantine the identified tests.
%	Quarantined tests do not get run as a part of your builds. They are there so that they can be fixed later.}
%\end{quote}

There are two strategies to deal with flakiness. Firstly, a practical approach to deal with flakiness as it occurs in a build is to trigger appropriate processes~\cite{mvn-gradle-flaky, gitlab-flaky}.  Failing tests are observed and then re-evaluated, and if they eventually pass, they are then marked as \textit{``success, but flaky''}. Projects can then respond to this by tagging those tests, quarantining them and/or automatically opening issues.  The focus here is on the current build cycle.
The second approach aims at  finding and addressing the root causes of flakiness to deal with it \textit{a priori}. Some of these being test order dependencies, concurrency, randomness, sensitivity to the test environment configuration, and dependency on network resources  \cite{parry2021survey}.
Detection usually utilises program analysis to inform refactoring~\cite{bell2015efficient,gambi2018practical,gyori2015reliable}.  %There are multiple techniques for detection that target some of these causes \cite{zhang2014empirically,gyori2016nondex,gambi2018practical,Lam2019Root,Shi2019iFixFlakies,lam2019idflakies,Mor2020FlakyLoc,Silva2020Shake,endo2020noderacer,Dong2021Concurrency}.

%Those two  approaches reflect different motivations: the process focus of industry that prioritises the need for a current build to succeed, and the focus of academic research that tries to understand the cause of the issue without always fully taking into account the cost to an organisation to then address it.
%\todo[]{It is better if we discuss the use in the contex of muliple languages runnin on JVM later when we proposed the new appraoch}
%Jens: I couldn't think about a better way, so kept it as is

We propose a novel approach to quarantine flaky tests in Java programs on-the-fly to make builds succeed without the need to rerun or rebuild. Our approach is based on a dynamic analysis inferring the assumption under which a failed test might pass, then disabling the test while also identifying the possible reason for flakiness. Analysis results are communicated in a way which allows processes or builds to proceed without requiring refactoring. This  retains the advantage that cause detection has in terms of provenance, i.e., our approach is not black box, it produces information as to why a test was skipped to assist engineers to address the sources of flakiness in the long run by informing the refactoring of code, tests and configuration scripts. There are multiple use cases for this general approach for sanitising flaky tests due to root causes such as network dependency, test order dependence and environment dependency.  For instance, tests could be disabled if (instrumented) test runs detect certain configuration (JVM versions, hardware, OS etc) either directly known or inferred to lead to flaky behaviour, or if executions of tests interfering with state a test relies on have been recorded.

We specifically illustrate the utility of our approach to deal with network dependencies in tests. This has been identified as one of the major sources of test flakiness~\cite{luo2014empirical,eck2019understanding,parry2021survey} and Luo et al.~\cite{luo2014empirical} classify 6\% of the flaky tests fixes they studied to be due to network dependencies. Our approach is particularly suitable here, as it is difficult or even impossible for engineers to write test fixtures or assumptions to reliably control or check the availability of networked resources when tests are run.

% More precisely, we investigate testing for programs that are compiled to run on the JVM, but are not necessarily written in Java~\footnote{In particular, one program in our dataset used for evaluation, \textit{jmeter}, contains a significant amount of \textit{groovy} code.}.

% Jens > Amjed : please more more detailed discussion also referencing some material from V-A into a new related work subsection.
%Network dependency (inclduing connections, availability, and bandwidth) has been acknowledged as one of the key factors that may lead to test flakiness \cite{luo2014empirical,eck2019understanding,parry2021survey}.
%This includes both local and remote network issues. Local issues pertain to managing resources such as sockets (e.g. contention with other programs for ports that are hard-coded in tests), while remote issues concern failures in connecting to remote resources.
%In a study of Python projects, Gruber et al. \cite{gruber2021empirical} noted that 413 tests are flaky due to network issues. It is also reported that 14\% of the flaky tests found in six large-scale Microsoft projects are due to network related causes \cite{lam2020study}. In a study consisting of Android projects \cite{Thorve2018empirical}, network is identified as a cause of flakiness of 8\% of the detected flaky tests.

%\todo[inline]{The introduction is missing a mention/discussion of our focus on network flakiness. I have added a smell paragraph to address issue \#7}
% Jens: done

This paper is organised as follows. We discuss the background in Section~\ref{sec:background} and develop the conceptual foundations for our approach in Section~\ref{sec:approach}. We then discuss the methods and tools we have developed in Section~\ref{sec:santisers}, followed by a discussion of how to apply them to deal with network dependencies in Section~\ref{sec:network-dependencies}. Section~\ref{sec:evaluation} contains the evaluation of our approach on a set of real-world programs with known  flakiness issues. This is followed by a discussion of related work in Section~\ref{sec:relatedwork} and a brief conclusion.
\\
We have included the scripts, datasets and results referenced in this paper in a replication package: \url{https://bitbucket.org/unshorn/flaky-test-sanitisation/}.
 % next paragraph is perhaps too cocky
%We underpin our approach by revisiting some fundamental concepts of testing, resulting in a notion of flakiness that better reflects the use of tests in automated processes than the currently used notion.

\section{Background}
\label{sec:background}

\subsection{Assumptions}
\label{ssec:background:assumptions}

A unit test (written in \textit{JUnit}) has a simple structure containing some computation, and some assertions to check the results of this computation against some expectations (oracles). A simple test is shown in Listing~\ref{code:simple}.

\begin{lstlisting}[language=Java, caption=A simple \textit{JUnit} test, label=code:simple]
@Test public void test() {
	int result = calculator.divide(4,2);
	assertEquals(2,result);
}
\end{lstlisting}

 The lifecycle of classes / objects to be tested (such as \texttt{calculator}) is usually managed outside the actual tests in a test \textit{fixture}. For instance, in \textit{JUnit}, those are methods annotated with \texttt{@BeforeEach}, \texttt{@BeforeAll}, \texttt{@AfterEach} and \texttt{@AfterAll}, respectively~\footnote{Throughout the paper, when we refer to \textit{JUnit} we mean \textit{JUnit5} unless explicitly mentioned otherwise. Most concepts discussed have counterparts in \textit{JUnit4} and older versions and are also present in alternative frameworks for automated unit testing.}.  A \textit{test run} is  a function mapping a set of tests to a \textit{success} or \textit{failure} state, i.e., $\rho : T \rightarrow \{success,failure\}$.  Flakiness occurs if there are two runs $\rho_1$ and $\rho_2$ producing different results for at least one test $t \in T$, i.e., $\rho_1(t) \neq \rho_2(t)$~\footnote{This nondeterminism could also be modelled by using a second \textit{environment} parameter, as done in~\cite{zhang2014empirically}. We chose to consider two runs as different functions for the sake of a compact presentation.}.

%Flakiness suggests some degree of non-determinism between runs. I.e., when considered as mathematical functions, two runs $\rho_1$ and $\rho_2$ are not necessarily the same function~\footnote{This could also be modelled via a second parameter $\rho : T \times E \rightarrow \{success,failure\}$ representing the environment, i.e. state that influences the evaluation of tests}.

Many testing frameworks, including \textit{JUnit}, define an additional state to describe the outcome of executing a test -- \textit{error}.  For instance, if the \texttt{divide} function used in Listing~\ref{code:simple} was invoked with a second parameter 0 in a test, then the test might result in an exception or error. Unless a test is specifically setup to test for this,  \textit{JUnit} would flag this outcome as \textit{error},  handling it slightly differently from \textit{failure}.  This raises the question whether a test suite should still be considered as flaky if between any two runs, the same tests pass (\textit{success}), but the execution of the tests sometimes results in \textit{failure}, and sometimes in \textit{error}. Failure and error are similar in that they reflect an unexpected state reached by the execution of test. The difference is merely whether this state is made \textit{explicit} (leading to \textit{failure}), or \textit{implicit} (leading to \textit{error}). The latter means that there is an implicit oracle \cite{barr2014oracle} that the execution of the functionality tested \textit{should not} result in an unhandled exception or error.

There is a fourth state that results from violation of \textit{assumptions}.  Consider the test shown in Listing~\ref{code:assumption}, and assume that the calculator to be tested is backed by a web service. Then the test is only meaningful if this service is reachable, and an assumption is used to express this.

\begin{lstlisting}[language=Java, caption=A JUnit test with assumptions, label=code:assumption]
@Test public void test() {
	assumeTrue(networkIsAvailable());
	int result = calculator.divide(4,2);
	assertEquals(2,result);
}
\end{lstlisting}

 Assumptions are not a widely utilised feature. Table~\ref {tab:junit-feature-use} demonstrates this -- it shows the number of test classes~\footnote{This refers to the number of \texttt{.class} files found containing  methods annotated as tests, methods with \texttt{@Test} or \texttt{@ParameterizedTest} annotations}, used standard assertions  (\textit{assrt.} column) and assumptions (\textit{assum.} column)~\footnote{Standard here refers to the assumptions and assertions defined as static methods in \texttt{org.junit.jupiter.api.Assumptions }(\textit{JUnit5}), \texttt{org.junit.Assume }(\textit{JUnit4}), \texttt{org.junit.jupiter.api.\-Assertions} (\textit{JUnit5}) and \texttt{org.junit.Assert} (\textit{JUnit4}) respectively, we are counting the call sites for those methods in the respective classes.} used and ignored or disabled tests (in the \textit{ign./dis.} column, these can be considered as the strongest assumptions that always fail), for both \textit{JUnit4} and \textit{JUnit5} APIs, in the programs used in the evaluation in Section~\ref{sec:evaluation}. Those features were extracted using an ASM-based bytecode analysis~\cite{bruneton2002asm} on the compiled test classes.

  % note: must run script in thsi folder to update table
  \begin{table}[]
\caption{Use of testing features in programs}
\label{tab:junit-feature-use}
\begin{tabular}{|lrrrrrr|}
\hline
 program  & classes   &  tests  & assrt. & assum.  & ign./dis. & cond. \\ \hline
biojava & 283 & 1,325 & 4,473 & 7 & 30 & 0 \\
jabref & 457 & 3,442 & 5,422 & 4 & 20 & 0 \\
jmeter & 222 & 1,584 & 3,290 & 5 & 5 & 4 \\
jsoup & 47 & 928 & 2,848 & 0 & 4 & 0 \\
pdfbox & 205 & 930 & 3,202 & 4 & 2 & 0 \\
swagger-parser & 1 & 25 & 0 & 0 & 0 & 0 \\
\hline\end{tabular}
\end{table}

\begin{lstlisting}[language=Java, caption=\textit{pdfbox} test with assumption, label=code:pdfbox-assumption]
@Test void isNonHeadlessPDFBoxTest() {
  final String[] args = {"debug"};
  assumeFalse(GraphicsEnvironment.isHeadless(), ..);
  assertDoesNotThrow(() -> {PDFBox.main(args);});
  ..
}
\end{lstlisting}

An interesting use of assumptions aligned with the approach presented here can be found in the tools module of \textit{pdfbox} (class \texttt{..PDFBoxNonHeadlessTest}), shown in Listing~\ref{code:pdfbox-assumption}. This  test is only meaningful if it is executed in a non-headless mode as a graphical user interface is created. This is often not the case when the test is evaluated within a container as part of a CI process. Nevertheless, the test is still useful when executed on a developer desktop.

\textit{JUnit5} introduced conditional tests using additional annotations to facilitate writing assumptions using common use cases, such as dependencies on the OS (\texttt{@DisabledOnOs}, \texttt{@EnabledOnOs}) or  the JRE (\texttt{@Dis\-abledOnJre}, \texttt{@EnabledOnJre}) used. \textit{Spock}~\footnote{\url{https://spockframework.org/}} has the \texttt{@Required} and \texttt{@IgnoreIf} annotations that take an expression (closure) as a parameter for a similar purpose. Occurrences of those features in the programs studied are listed in the \textit{cond.} column in Table~\ref {tab:junit-feature-use}. %Again, this is an underutilised feature. @Amjed: I have comments this out as it is redundant. Also Spock was used only in jmeter

 The semantics of assumptions is defined in the \textit{JUnit} documentation as follows:

  \begin{quote}
\textit{A failed assumption does not mean the code is broken, but that the test provides no useful information. Assume basically means ``don't run this test if these conditions don't apply''. The default JUnit runner skips tests with failing assumptions. ~\footnote{\url{https://junit.org/junit4/javadoc/4.12/org/junit/Assume.html}}.}
  \end{quote}

That is, a dedicated \textit{skip} state is used by test runners to mark tests failing assumptions ~\footnote{There are other reasonable interpretations of failed assumptions, for instance, a test could be interpreted using the semantics of the material implication, meaning that a test would succeed if its assumptions failed. The introduction of another state however has the advantage of adding some provenance to testing.}.

The duality of assumptions and assertions stipulates that tests follow the general pattern of a specification, often represented by the triples in Hoare-Floyd logic~\cite{hoare1969axiomatic}, with assumptions corresponding to preconditions, and assertions to postconditions. This relationship is well-known and has been exploited, for instance, to generate tests from formal specifications~\cite{cheon2002simple,boyapati2002korat}.
However, there are some important differences that are relevant for our discussion here: in Hoare-Floyd logic there is no dedicated state to express failed preconditions, whereas tests use a dedicated multivalued logic to deal with failed assumptions with a dedicated \textit{skip} (\textit{unknown}) state. What is more, tests cannot be assumed to be side-effect-free functions, they often modify memory or external resources. Finally, some tests are inherently concurrent even  if  they are not executed concurrently. An example for this is discussed later in Section~\ref{sec:network-dependencies}, where a test execution relies on network connectivity provided by an OS process which cannot be  controlled by a fixture, or even on the availability of DNS, routers and servers. There is no counterpart for this in Hoare-Floyd logic, although there are extensions to model concurrency~\cite{jones1983specification,reynolds2002separation}.

\subsection{Notions of Flakiness}
\label{ssec:background:notions-of-flakiness}

Usually, flakiness is defined as non-deterministic test outcome where the same test sometimes succeeds and sometimes fails (which is signalled by either the \textit{failure} or the \textit{error} state). By also taking the \textit{skip} state into account, we may also consider a weaker notion of flakiness. \textit{Weak flakiness} occurs if the outcome of any test evaluated changes. In particular, this includes changes of test outcomes between runs from \textit{success} to \textit{skip}, and vice versa.

When considering the default configuration of many build systems, the notion of \textit{weak flakiness} may not be very interesting, as it does not carry the risk of builds failing. For instance, consider the most widely used Java build tool: Apache Maven~\footnote{\url{https://www.jetbrains.com/lp/devecosystem-2021/java/\#Java_which-build-systems-do-you-regularly-use-if-any}}, with tests being executed using the test lifecycle phase with the default \textit{surefire} configuration. While any test resulting in a \textit{failure} or \textit{error} state will lead to a build failure, tests resulting in a \textit{skip} state do not, thus builds can proceed.

On the other hand, if weak flakiness is detected and a respective signal is emitted, processes can be customised to respond to this, for instance, by creating issues.  This aligns with the idea of emitting orange signals (in addition to green / red) in order to break free of the `\textit{`boolean straight jacket'}' assumption~\cite{bojarczuk2021measurement}.

\subsection{Explicit vs Implicit Assumptions}
\label{ssec:background:explicit-vs-implicit-assumptions}

The impact of evaluating  assumptions and assertions on state can be summarised in Table~\ref{tab:test-state}.

\begin{table}[h]
\caption{Test result states in JUnit}
\center
	\begin{tabular}{|llll|}
		\hline
		condition  & explicit & on success                  & on failure                  \\ \hline
		assumptions & yes      & continue                    &  \textit{skip}    \\
		% & no       & ?                           & ?                           \\
		assertions  & yes      & \textit{success} &  \textit{failure} \\
		& no       &   \textit{success} & \textit{error}  \\ \hline
	\end{tabular}
\label{tab:test-state}
\end{table}

What is noticeable is that there is no provision for implicit assumptions. The (meta-)assumption is that there is no need for this, as all assumptions can be made explicit.
%. Not having assumptions for a test is the equivalent of the weakest possible precondition that is always satisfied.
This hinges on the expectation that an engineer can completely control and check the program and the environment it interacts with. We believe that this expectation is flawed  for two reasons: (1) many parameters a test execution relies on cannot be controlled and not even be checked and (2) even if this was possible, this would require an unrealistic level of  understanding of the program and its interactions with the environment  by the engineer(s) in charge of testing.

However, these insights also provide the opportunity to explore the very notion of implicit assumptions further, and investigate whether this can lead to a novel approach to tackle certain kinds of flakiness. 
% Jens: disabled next sentence, too repetetive 
%We explore this question further in the rest of this paper.

% Jens: I took this out, the new preceeding paragraph is more descriptive
%\begin{enumerate}
%	\item The rise of multi-core CPUs where process and thread scheduling depends on hardware, operating systems, and has some inherent non-determinism that cannot be controlled by an application.
%	\item The popularity of continuous integration where code development and  test execution run on different computers, with subtle differences.
%	\item The quick software update cycles where small optimisations can have observable effects in programs (example: hashcode optimisations and iteration order of hashed containers), changing the behaviour of code programmers may consider stable.
%	\item Cloud-based deployment that uses abstractions (like docker) from actual hardware and operating systems, blurring dependencies without completely removing them.
%	\item The interaction of programs with networked services.
%\end{enumerate}
%Many of those aspects can not be (easily) controlled with a test fixture, so other means are required to address them as potential sources of flakiness.

\section{Strengthening Assumptions as Deflaking Strategy}
\label{sec:approach}

\subsection{Deflaking Strategies}
\label{ssec:approach:strategies}

Given the structure of a test, there are multiple ways to tackle flakiness. This is very much a ``horses for courses'' scenario, i.e., different approaches might be suitable for different situations. An empirical study by Luo et al. \cite{luo2014empirical} has shown several methods are used by developers to address flakiness.  Possible strategies include:

\begin{itemize}
	\item \textbf{Modifying the Program under Test} in order to improve testability ~\cite{ISO25010},  examples include avoiding the unnecessary use of concurrency and randomness, and controlling aliasing in order to reduce dependencies between tests.
	\item \textbf{Modifying the Test Fixture}, examples include cleaning up resources (heap or external) used by tests that may lead to test dependencies, or setting random seeds.
	\item \textbf{Reconfiguring the Test Environment}, examples include using fork options in tools like Maven surefire, and using a customised JVM to better control test dependencies~\cite{bell2014unit}).
	\item \textbf{Weakening Assertions}, examples include the introduction of deltas when comparing numerical values, and replacing fixed oracles with statistical oracles~\cite{mayer2004test,hewson2015performance}.
\end{itemize}

\subsection{Strengthening Assumptions}
\label{ssec:approach:strengthenassumptions}

There is another option to deal with flakiness: the strengthening of assumptions. Given that we can interpret no assumption as the weakest possible assumption that always evaluates to true, this usually means adding assumptions to tests.
%Jens: commented out  to save space
%As demonstrated above,  assumptions are not widely used by developers.
%I.e. developers often only use the weakest possible implicit assumption (no assumption) that always evaluates to true. Interestingly, when developers do use assumptions, they then tend to use the strongest possible assumption that always evaluates to false, using the special annotations \texttt{@Ignore} and \texttt{@Disabled}.

Strengthening assumptions is particularly suitable to state dependencies on relevant aspects of the test environment that cannot be controlled in fixtures.
%They can also serve as contracts between the test engineers writing the tests, and the devops engineers defining the environment used for testing.

%\begin{enumerate}
%	\item Not all aspects of a test environment can be controlled, for instance, if a test relies on a network connection to an external service that cannot be effectively mocked.
%	\item If the test environment is controlled, then this is usually done by a devops engineer, not by the engineer writing the tests. Assumptions can serve as contract between those stakeholders.
%	\item Setting up a test environment is a potentially expensive task, while inferring assumptions can sometimes be automated as we demonstrate in this paper.
%\end{enumerate}

\subsection{Static vs Dynamic vs External Assumptions}
\label{ssec:approach:staticvsdynamic}

Looking at assumptions, we can broadly classify them as follows: \textbf{Static assumptions} that  always evaluate to \textit{true} or \textit{false} --  this includes the empty assumption always evaluating to true and the strongest possible assumption defined by  the special annotations like  \texttt{@Ignore} or \texttt{@Disable} that can never be satisfied.
%The static nature allows testing frameworks to perform optimisations, for instance, to skip the execution of fixtures for disabled tests.
On the other hand, \textbf{dynamic assumptions}  are evaluated before the actual test is executed, but the result of the evaluation may be different for each execution. This is the kind of assumption supported by the assumption APIs (\texttt{Assume} and \texttt{Assumptions}, respectively) in \textit{JUnit}.
Finally, there are \textbf{external assumptions} -- assumptions related to the execution environment of the test that cannot be checked using a unit testing API. However, those assumptions may still emit observable effects and therefore actionable signals during test execution (i.e., they require a \textit{speculative execution} of the test to be observed). The main use case is the presence of reliable network connections during testing that cannot be checked or controlled via some testing API, but has a direct impact on the  outcome of tests. This will be discussed in detail in Section~\ref{sec:network-dependencies}.

The rest of this paper focuses on external assumptions. Support for external assumptions is provided by means of \textit{sanitisers}, testing framework extensions that intercept test processing, and can check inferred assumptions on-the-fly, i.e., without an explicit representation of these assumptions in tests. This means that flaky tests that would otherwise result in \textit{failure} or \textit{error} are skipped, allowing builds to proceed (but keeping a record of those tests).

% jens: disabled to manage space, already discussed in related work now
%The main use case for this is to deal with flakiness caused by unstable network connections, we will discuss this in detail in Section~\ref{sec:network-dependencies}. However, the applicability of the idea is not restricted to network dependencies. There is another use case with existing support in  \textit{spock}~\footnote{\url{https://spockframework.org/}} and \textit{groovy}~\footnote{\url{http://docs.groovy-lang.org/}} that at least partially aligns with the approach presented here. \textit{Spock's} \texttt{@PendingFeature} and \textit{groovy's} \texttt{@NotYetImplemented} annotations are used to intercept and changed test results. These annotations can be interpreted as assumptions that the code tested is not completely implemented. In particular, \texttt{@PendingFeature} will mark failed annotated tests as \textit{skipped} (so build can still complete). Interestingly, it will flag passing annotated tests as \textit{error} to signal that the annotation should now be removed.

%Jens: on second thought this might be misleading
%In terms of the terminology proposed earlier, the use of santitisers can transform test suits that are strongly flaky into test suites that are ``just''  flaky.

\section{Sanitisers}
\label{sec:santisers}

\subsection{Sanitisers as JUnit5 Extensions}
\label{ssec:santisers:junit5}

We have developed a proof-of-concept implementation to sanitise~\footnote{We use sanitise in the sense of ``to make (something, such as text) more acceptable by removing, hiding, or minimizing any unpleasant, undesirable, or unfavourable parts' [\url{https://www.merriam-webster.com/dictionary/sanitize}]} flaky tests called \textit{saflate}. \textit{Saflate} sanitisers intercept  the execution of tests, inspect the state of tests, checks inferred assumptions and depending on those conditions will change the state of some tests from \textit{failure} or \textit{error} to \textit{skip}.  In \textit{JUnit5}, states are set and propagated via special errors: \texttt{java.lang.AssertionError} are used to signal failed tests, whereas \texttt{org.opentest4j.Incomplete\-Execution\-Exception} and its subclasses are used to signal skipped and aborted~\footnote{Aborted means a failed assumption, whereas skipped means ignored due to an annotation. We refer to both situations as skipped throughout this paper.} tests. Changing the test state is therefore a matter of intercepting test processing, catching certain exceptions or errors, and rethrowing others. 

\textit{Saflate} is implemented using the \textit{JUnit5} extension model~\cite[Sect 5]{junit5userguide}, it requires \textit{JUnit 5.6.0} or better. Extensions implement some test framework interfaces, and be deployed programmatically, declaratively using annotations, or automatically as services. The annotation-based approach is class-based, for instance, the code in Listing~\ref{code:extension} will install the network dependency annotation extension  discussed in Section~\ref{sec:network-dependencies} for the tests within the class \texttt{MyTests}.

\begin{lstlisting}[language=Java, caption=Using the network dependency sanitiser extension, label=code:extension]
import .. .SanitiseNetworkDependenciesExtension;
import org.junit.jupiter.api.extension.ExtendWith;

@ExtendWith(SanitiseNetworkDependenciesExtension.class)
public class MyTests {
	@Test public void test() {
		..
	}
}
\end{lstlisting}
 
Using the automated service-based mechanism, developers have to add  \textit{saflate} as a dependency to the project, without the need for code refactoring. \textit{JUnit5} will then automatically discover the service and inject it into the test processing pipeline. The only additional action required is to  set the \texttt{junit.jupiter.ex\-ten\-sions.auto\-detection.enabled} system property to \texttt{true}, this can be done via a JVM argument~\footnote{For instance, in Maven, this can be configured in the \textit{surefire} plugin used for testing}.
 
Some sanitisers need to instrument existing classes. For the sanitisers discussed here, this is done dynamically using \textit{ByteBuddy}~\footnote{\url{https://bytebuddy.net/}} without the need of installing an agent.  The instrumentation is triggered by the static blocks of extension classes. 
% Jens: commented out to save some space
%An example where this approach is used is  the instrumentation of some  exception classes in the standard library used to detect network dependencies in test. This will be discussed in Section~\ref{sec:network-dependencies}.

The semantics of the sanitiser is defined by the following two rules:

\begin{enumerate}
	%\item If the test state is \textit{success} then it proceeds uninterrupted.
	%\item If the test state is \textit{skip} then the test proceeds uninterrupted.
	\item[R1] If the test state is \textit{failure} or \textit{error},  and an inferred assumption is not satisfied, then its state is set to \textit{skip}, and the test proceeds. 
	\item[R2] Otherwise, the test proceeds uninterrupted.
\end{enumerate}

When the state is changed to \textit{skip}, an instance of the respective exception that signals the skip state within the respective testing framework is created and thrown. This facilitates \textit{provenance} -- sanitisers add a descriptive message as to why the test was skipped that is then included in generated reports and communicated to users, where it can inform refactoring. 

Multiple such extensions can be combined by chaining the respective extensions. If multiple preconditions associated with extensions are violated, only the first one encountered will change the state of the test to \textit{skip}, as tests with a state \textit{success} and \textit{skip} are not intercepted.  This has some impact on provenance, which is provided through the messages of the exceptions signalling that an assumption was not satisfied: only the details about the first violated assumption are signalled. When extensions are deployed programmatically or using annotations, the user has control over the order in which extensions are being processed~\cite[Sect 5]{junit5userguide}.

 %  (i.e. it resulted in a \texttt{java.lang.AssertionError})  \texttt{org.open\-test4j.Incomplete\-ExecutionException} \texttt{org.opentest4j.TestAbortedException}
 
 \subsection{Dealing with Test Fixtures}
 \label{ssec:santisers:fixtures}
 
 The extensions consider the \texttt{@BeforeEach} and \texttt{@AfterEach} fixtures as part of the test. This is consistent with standard \textit{JUnit5} behaviour -- if either of these methods results in an \textit{error}, so will the respective tests. However, \texttt{@BeforeAll }and \texttt{@AfterAll }are not intercepted -- errors occurring in those methods lead to all tests in scope being skipped.
 
 \subsection{Backporting Extensions as JUnit4 Rules}
 \label{ssec:santisers:junit4}
 
 Some programs we used to evaluate our approach employ the older \textit{JUnit4}. For this purpose, we backported saflate extensions as \textit{JUnit4} rules.  In order to deploy those rules,  tests have to be modified by inserting those rules as fields since \textit{JUnit4} does not support the service-based auto-deployment available for \textit{JUnit5} extensions.

 \subsection{Implementing Sanitisers as Spock Extensions}
  \label{ssec:santisers:spock}
  
 \textit{Spock}~\cite{kapelonis2016java} is an increasingly popular testing framework for Java and Groovy, written in those languages.  It supports an annotation-based extension mechanism. The semantics of an annotation extending \textit{spock} can be provided by an \textit{interceptor} that intercepts the testing pipeline and can change the state of a test. We followed the overall designing of built-in \texttt{@PendingFeature} annotation discussed earlier and again in Section~\ref{ssec:relatedwork:inthewild}.

\section{Application: Sanitising Network Dependencies}
\label{sec:network-dependencies}

\subsection{Motivation}
\label{ssec:network-dependencies:motivation}

Modern applications are increasingly networked as remote services and resources are both provided and consumed. This causes challenges for testing as tests rely on a stable network connection to succeed. Problems related to network connectivity have recently been identified as a major cause for failing and restarted builds on Travis CI~\cite{durieux2020empirical}, and in organisations like Uber~\cite{Agarwal2021Handling} and Mozilla~\cite{lampel2021life}.

The provision of stable network connections can generally not be ensured in a test fixture. There is no reliable way for engineers to write assumptions that certain tests are only executed
when the network is available~\footnote{This topic frequently comes up in discussions, for instance see~\url{https://stackoverflow.com/questions/28397007/how-to-simulate-lack-of-network-connectivity-in-unit-testing}.}.  And even if there was such an API, it would be of limited use, as network connectivity could be lost when a test is executed, but after the respective assumption has been checked.

Many programmers prefer to mock APIs and services relying on network connectivity using approaches such as dynamic mock libraries (\textit{mockito}\footnote{\url{https://site.mockito.org/}} etc), static network-specific libraries such as Springs servlet API mock objects~\footnote{\url{https://docs.spring.io/spring-framework/docs/current/reference/html/testing.html\#mock-objects}}, or more complex frameworks based on  service virtualisation \cite{versteeg2016opaque}. However, tests directly relying on resources accessed via the network are still common. This makes network dependencies a major source of flakiness~\cite{luo2014empirical,eck2019understanding,parry2021survey}.

\subsection{An Optimistic Approach: Hoping for the Best}
\label{ssec:network-dependencies:optimistic}

Consider the  test\footnote{\url{https://github.com/JabRef/jabref/blob/bb011c9313367a28990ae213b3920fe6cd10d1dc/src/test/java/org/jabref/logic/help/HelpFileTest.java}} shown in Listing \ref{code:jabref}, taken from the popular \textit{jabref} project - an application to manage bibtex databases.

% \lstinputlisting[language=Java, caption=PDFBox test, label=code:pdfbox]{code/pdfbox.java}

\begin{lstlisting}[language=Java, caption=A \textit{jabref} test with a network dependency, label=code:jabref]
package org.jabref.logic.help;
..
class HelpFileTest {
  private final String jabrefHelp = "https://docs.jabref.org/";
  static Stream<HelpFile> getAllHelpFiles() {
    return Arrays.stream(HelpFile.values());
  }
  @ParameterizedTest
  void referToValidPage(HelpFile help) throws IOException {
    URL url = new URL(jabrefHelp + help.getPageName());
    HttpURLConnection http = 
      (HttpURLConnection) url.openConnection();
    http.setRequestProperty("User-Agent",URLDownload.USER_AGENT);
    assertEquals(200, http.getResponseCode(), ..);
  }
}
\end{lstlisting}

%  \todo[inline]{Jens: perhaps org.jabref.logic.help.HelpFileTest is a better example}

The help system uses online resources, and this functionality is tested here. From the project's point of view, this has merits: an online help system is easier to maintain, and this avoids adding (potentially large binary) resources to the repository and to the distribution. But if those tests fail due to the network not being available then this does not indicate that the code is incorrect, it merely states that these particular tests cannot be evaluated at the current time, and therefore should be skipped. Overall, the approach taken here is \textit{optimistic}, in the sense that the developers expect the network to work and the site to be available and reachable.

\subsection{A Pessimistic Approach: Reducing Coverage (but ``not reliant on the vagaries of the internet'')}
\label{ssec:network-dependencies:pessimistic}

Now consider a following test\footnote{\url{https://github.com/jhy/jsoup/blob/master/src/test/java/org/jsoup/integration/UrlConnectTest.java}} (Listing~\ref{code:jsoup}) in \textit{jsoup} - a popular HTML parser library. Here, developers took a \textit{pessimistic} approach by disabling network-sensitive tests. While this reduces flakiness caused by network problems, it also unnecessarily removes valuable tests when the network is available as the assumption is static.

\begin{lstlisting}[language=Java, caption=\textit{jsoup} test with network dependency, label=code:jsoup]
package org.jsoup.integration;
..
/**Tests the URL connection. Not enabled by default, so 
 tests don't require network connection. .. */
@Disabled // ignored by default so tests don't require network access. comment out to enable. todo: rebuild these into a local Jetty test server, so not reliant on the vagaries of the internet.
public class UrlConnectTest {
  ..
  @Test public void fetchBaidu() throws IOException {
    Connection.Response res = Jsoup
      .connect("http://www.baidu.com/")
      .timeout(10*1000).execute();
    Document doc = res.parse();
    ..
\end{lstlisting}

  The two examples illustrate that developers often do not have good choices -- either test less, or accept flakiness which leads to failing tests. This is the problem we are trying to address in this work.

\subsection{Hidden Network Dependencies}
\label{ssec:network-dependencies:hidden}

Network dependencies can be subtle, and the engineer writing tests may not always be aware of their presence. For instance, consider the case of XML parsing. Often, XML documents reference some remote schema (XSD, DTD or similar) using a URL. Many applications use validating parsers to check that documents comply to a vocabulary. This requires network access at the time of validation. Even if the parser validation is disabled, schema access may still be required to resolve entity references. Other subtle causes of network-related flakiness are (tight) connection timeouts and port collisions. %This has been identified as a major source of flakiness at Uber~\cite{Agarwal2021Handling}.

\subsection{Implementation}
\label{ssec:network-dependencies:implementation}

The condition enforced by the sanitiser is that no exception indicating a network problem must have occurred during the execution of the test. We consider three such exceptions (including subclasses of those listed) related to network availability:
\begin{enumerate}
	\item \texttt{java.net.SocketException}
	\item \texttt{java.net.UnknownHostException}
	\item \texttt{java.net.NoRouteToHostException}
\end{enumerate}

We implemented a sanitiser (called \textit{saflate}) for network dependencies as  both a \textit{JUnit5} extension and a \textit{JUnit4} rule, and an experimental port as a \textit{Spock} extension. The list of exceptions to be sanitised is provided by a service, allowing the list to be extended by third parties. The standard Java \texttt{ServiceLoader} facility is used for this purpose, i.e. extension can be provided as libraries that advertise the implementation of the \texttt{NetworkExceptionProvider} interface in the component manifest. \textit{saflate} source code is available on a public GitHub repository \footnote{\url{https://github.com/jensdietrich/saflate}}, and the deployed binaries are available in the Maven repository \footnote{\url{https://search.maven.org/search?q=io.github.jensdietrich.saflate}}

The sanitisers instrument the constructors of the respective exception classes.
When the creation of such an exception is encountered, a flag is set to communicate to the test that a network exception has occurred. When a test fails or results in an error, then the test state is changed to \textit{skip} if either (1)  the test resulted directly in a network exception visible in the stack trace and the test is in an \textit{error} state or (2) a network exception creation was recorded during  test execution, and the test is in a \textit{failure} state.

Communicating  the state from the network exception instantiation to tests has to deal with classloader issues as those exceptions are loaded by the system classloader. We solved this problem by using a system property with a dedicated name in order to communicate properties between those exceptions, and the customised test execution. The property can also incorporate the thread ID in case tests are executed using thread-level concurrency~\footnote{This behaviour is by default disabled, but can be enabled by setting the system property \texttt{saflate.supportconcurrent-test-execution} to \texttt{true}}. This is an experimental feature introduced in \textit{JUnit5}, but is used by both \textit{pdfbox} and \textit{jabref} in order to improve test performance.
% reference broken if pdfbox example is not used
%The respective annotation can be seed in line 3 in the test shown in Listing~\ref{code:pdfbox}.

Note that the condition we enforce is  not a precondition as we can only detect a violation after the test has been at least partially executed, i.e., \textit{a posteriori}. We can consider the execution of the test as speculative execution if a precondition holds, but interrupting a test if we find out later that this was not the case, relying on a fixture to rollback, i.e., to reset the state of the test.  A clean precondition check would be certainly preferable, however, there is no way to check this, and then also to ensure network connectivity as a guarantee during the execution of the test.

\subsection{Limitations}
\label{ssec:network-dependencies:limitations}

% Jens > Amjed -- I decided to leave it here, as this sets the scene for measuring precicion and recall discussed next
%\todo[]{I'd suggest to move this to the end of the results section, or even to the end of the paper}

This approach has the following limitations that may result in false positives (unnecessarily sanitised tests) and false negatives (tests that should be sanitised, but are not).

\textit{Saflate} will suffer from false positives, if network exceptions are created but not actually being thrown or intentionally thrown to test the exceptions (e.g. using \texttt{assertThrows}), but the test nevertheless results in \textit{error} or \textit{failure} for some other  reason. We consider this scenario as very unlikely and suggest  mitigating this by using the more fine-grained annotation-based deployment if such tests are used.

On the other hand, \textit{saflate} will suffer from false negatives if support for  test concurrency is enabled and network exceptions occur in background threads.  This may result in tests not being sanitised.  A second potential source of false negatives are network exceptions thrown by native methods, such as \texttt{java.net.In\-et6AddressImpl::lookupAllHostAdd}, as the respective instantiations are not instrumented.  This would only result in a false negative if the respective test resulted in \textit{failure} (not \textit{error}).  It is possible to mitigate this by using a more sophisticated instrumentation that injects probes at the call / allocation sites of those constructors. This however is likely to increase the performance overhead.

The evaluation reports recall and precision in order to assess the impact those limitations have in practice. While we find instance of both false negatives and false positives, their relative numbers remain very low as we will demonstrate in the next Section.

\section{Evaluation}
\label{sec:evaluation}

\subsection{Dataset}
\label{ssec:evaluation:dataset}

Finding open-source projects with tests that experienced flaky behaviour due to network issues is challenging, as we noticed that such flaky behaviour might not appear until an interception to the network connection occurs if these tests (or the code being tested) rely on a network connection.

For our experiment, we constructed a dataset of projects with network dependencies from GitHub. We aimed to select only projects that are popular (with at least 300 stars) and with a decent history (over 2 years of age). Our goal is to select projects that are currently active and frequently maintained, thus we can report any flakiness we find to the maintainers.

We first searched for projects with open issues that are related to tests with flaky network behaviour. We used a simple search string \textit{``flaky + network"}, but this search, based on the results in the first 10 pages, did not retrieve any relevant results. We then searched for issues that included the specific network-related exceptions that we intercept:  \textit{``flaky + (UnknownHostException OR NoRouteToHostException OR SocketException)"}. We identified only one project with a current open issue with a network-related flaky behaviour (\textit{biojava}) that also met our selection criteria. We then decided to expand our search by conducting a general search on GitHub to identify tests that use specific HTTP GET requests such as  \texttt{java.net.URL::openconnection} and \texttt{java.net.URL::openStream}. Based on this, we identified five additional projects (\textit{jabref}, \textit{jsoup}, \textit{pdfbox} and \textit{swagger-parser} and \textit{jmeter}).
The six projects are listed in Table~\ref{tab:networkdependencies-dataset}~\footnote{For \textit{swagger-parser}, we only consider the \textit{swagger-parser} module. There are three other modules in this project, all using \textit{TestNG}, which does not offer suitable extension points to port \textit{saflate}, so we had to exclude these modules. The \textit{swagger-parser} module still uses \textit{TestNG} assertions, however, they are fully compatible with \textit{JUnit4}.}.

%We have evaluated this approach on five popular real-world projects with known network dependencies which we identified in a GitHub issue search, listed in Table~\ref{tab:networkdependencies-dataset}~\footnote{For \textit{swagger-parser}, we only consider the \textit{swagger-parser} module. There are three other modules in these project, all using TestNG. TestNG does not offer suitable extension points to port \textit{saflate}, so we had to exclude these modules. The \textit{swagger-parser} module still uses TestNG assertions, however, they are fully compatible with JUnit4.}.

\begin{table*}[]
\caption{Programs with tests depending on network connectivity}
\resizebox{\linewidth}{!}{
\begin{tabular}{|lllllll|}
\hline
program & description & tag & \begin{tabular}[c]{@{}l@{}}GitHub \\ stars\end{tabular} & \begin{tabular}[c]{@{}l@{}}testing \\ framework\end{tabular} & \begin{tabular}[c]{@{}l@{}}concurrent \\ tests\end{tabular} & repo url \\ \hline
biojava & a set of tools to process biological data & biojava-6.0.4 & 486 & junit 5 and 4 & no & https://github.com/biojava/biojava \\
jabref & citation and reference management tool & v5.5 & 2.5k & junit 5 & yes & https://github.com/JabRef/jabref \\
jmeter & performance and load testing tool & rel/v5.4.3 & 6k & \begin{tabular}[c]{@{}l@{}}junit 5 and 4 \\ spock\footnote{\url{https://spockframework.org/}}\end{tabular}  & no & https://github.com/apache/jmeter \\
jsoup & Java HTML parser & jsoup-1.14.2 & 9.4k & junit 4 & no & https://github.com/jhy/jsoup \\
pdfbox & tool for working with PDF documents & 2.0.24 & 9.4k & junit 5 & yes & https://github.com/apache/pdfbox \\
swagger-parser & parser for OpenAPI definitions  & v2.0.18 & 559 & junit 4 & no & https://github.com/swagger-api/swagger-parser\\
\hline
\end{tabular}}
\label{tab:networkdependencies-dataset}
\end{table*}

% \begin{table}[]
% 	\begin{tabular}{|llll|}
% 		\hline
% 		program        & tag & junit-version & concurrent tests \\ 	\hline
% 		biojava        & biojava-6.0.4    & 4             & no               \\
% 		jabref         & v5.5    & 5             & yes              \\
% 		jsoup          & jsoup-1.14.2    & 4             & no               \\
% 		pdfbox         & 2.0.24    & 5             & yes              \\
% 		swagger-parser & v2.0.18    & 4             & no      \\ 	\hline
% 	\end{tabular}
% 	\caption{Programs with tests depending on network connectivity.}
% \label{tab:networkdependencies-dataset}
% \end{table}

\subsection{Experimental Setup}
\label{ssec:evaluation:setup}

We run each program with four different configurations with and without the network enabled, and with and without the \textit{saflate} sanitiser. For \textit{pdfbox} and \textit{jabref}, we enabled \textit{saflate's} support for parallel test execution, discussed in Section \ref{ssec:network-dependencies:implementation}. Experiments were run in Docker containers that are made available in our replication package \footnote{\url{https://bitbucket.org/joe-bloggs/flaky-test-sanitisation/}}. The host used is a computer with a 3.2 GHz 6-Core Intel Core i7 CPU running Docker Engine - Community  20.10.5.

\subsection{Results}
\label{ssec:evaluation:results}

Table~\ref{tab:unsanitized} and Table~\ref{tab:sanitized} summarises the results obtained from the evaluation of the three programs~\footnote{Test counts appear to be inconsistent with Table~\ref{tab:junit-feature-use}. This is caused by parameterised tests -- the static analysis counts them as single tests, whereas test execution reports each instantiation separately.}. Table~\ref{tab:precrecall} lists the precision and recall scores for sanitisation.
The results reported are averages taken over 10 runs - numbers in brackets indicate that those are average numbers with some variation between runs.  We assess precision and recall as follows.  The \textit{sanitised tests }$t_s$  are the tests that result in failure or error in all runs when the network is off without sanitisation, but are always skipped when sanitisation is used and the network is off.   \textit{Relevant tests }$t_r$ are the test that always succeed when the network is available and sanitisation is off, but always result in failure or error when the network is off and no sanitisation is used. Precision is then defined as $|t_r \cap t_s| /| t_s| $, and recall as $|t_r \cap t_s| /| t_r| $. Multiple runs are used to deal with other sources of flakiness that may be present. These metrics  are based on the assumption that we can set up an experiment  that can control access to networked resources by disabling the network in the (docker-based) setup -- this may not work if the problem is not the connection, but the availability of the server. An indicator of this is when tests already fail when the network is available, and can be sanitised successfully.

The results indicate that generally \textit{saflate} is effective in sanitising tests failing or resulting in error due to network availability problems and poses no significant overhead. For \textit{jsoup}, \textit{pdfbox}, \textit{swagger-parser} and \textit{jmeter}, the results are perfect in the sense that \textit{saflate} sanitises exactly the tests that stop succeeding when the network is disabled. Interestingly, in case of \textit{swagger-parser}, \textit{biojava}, \textit{jabref} and \textit{jmeter}, disabling the network results in failures as opposed to errors. I.e., when intercepting test processing, the error stack trace containing references to the network connection may not be available. This justifies our choice to use a dual strategy -- instrument the allocation sites of network exceptions and analyse stack traces when available.
\begin{table*}[]
\caption{Test results without sanitisation. Results are reported across 10 test runs, numbers in brackets means that we observed some variation across runs and report the mean. The number of weakly flaky (w. flaky) and flaky  tests across those runs are reported as well.  Runtimes are reported in seconds (rt(s)).}
\label{tab:unsanitized}
	\begin{tabular}{|l|r|r|r|r|r|r|r|r|r|r|r|r|r|r|}
		\hline
		\multirow{3}{*}{program} & \multirow{3}{*}{tests} & \multicolumn{12}{c|}{unsanitised}  \\ \cline{3-14}
		&  & \multicolumn{6}{c|}{network on} & \multicolumn{6}{c|}{network off}  \\ \cline{3-14}
				& & failure  & error    & skipped & w. flaky & flaky & rt (s) & failed & error   & skipped  & w. flaky & flaky  & rt (s)   \\ \hline
biojava & 1,469 & 1 & 0 & 30 & 0 & 0 & (1102.3) & 17 & 281 & 33 & 2 & 2 & (117.2) \\
jabref & 7,586 & 1 & 0 & 19 & 0 & 0 & (173.2) & 57 & 0 & 19 & 0 & 0 & (108.3) \\
jmeter & 3,545 & 1 & 0 & 13 & 0 & 0 & (96.2) & 4 & 0 & 13 & 0 & 0 & (349.4) \\
jsoup & 930 & 0 & 0 & 3 & 0 & 0 & (3.0) & 0 & 1 & 3 & 0 & 0 & (2.8) \\
pdfbox & 1,881 & 0 & 0 & 2 & 0 & 0 & (82.2) & 0 & 35 & 2 & 0 & 0 & (67.5) \\
swagger-parser & 25 & 0 & 0 & 0 & 0 & 0 & (3.6) & 2 & 0 & 0 & 0 & 0 & (2.0) \\  \hline
	\end{tabular}
\end{table*}

\begin{table*}[]
\caption{Test results with sanitisation. Results are reported across 10 test runs, numbers in brackets means that we observed some variation across runs and report the mean. The number of weakly flaky (w. flaky) and flaky tests across those runs are reported as well.  Runtimes are reported in seconds (rt(s)).}
\label{tab:sanitized}
	\begin{tabular}{|l|r|r|r|r|r|r|r|r|r|r|r|r|r|r|}
		\hline
		\multirow{3}{*}{program} & \multirow{3}{*}{tests} & \multicolumn{12}{c|}{sanitised}  \\ \cline{3-14}
		&  & \multicolumn{6}{c|}{network on} & \multicolumn{6}{c|}{network off}  \\ \cline{3-14}
				& & failure  & error    & skipped & w. flaky & flaky & rt (s) & failed & error   & skipped  & w. flaky & flaky  & rt (s)   \\ \hline
biojava & 1,469 & 1 & 0 & (30.1) & 1 & 0 & (1228.1) & 1 & 7 & 323 & 2 & 0 & (227.7) \\
jabref & 7,586 & 0 & 0 & 20 & 0 & 0 & (190.0) & 0 & 0 & 76 & 0 & 0 & (109.5) \\
jmeter & 3,545 & 1 & 0 & 13 & 0 & 0 & (122.9) & 1 & 0 & 16 & 0 & 0 & (355.5) \\
jsoup & 930 & 0 & 0 & 3 & 0 & 0 & (3.3) & 0 & 0 & 4 & 0 & 0 & (3.1) \\
pdfbox & 1,881 & 0 & 0 & (2.1) & 1 & 0 & (85.5) & 0 & 0 & 37 & 0 & 0 & (70.3) \\
swagger-parser & 25 & 0 & 0 & 0 & 0 & 0 & (4.3) & 0 & 0 & 2 & 0 & 0 & (2.7) \\  \hline
	\end{tabular}
\end{table*}
\begin{table}[]
\caption{Precision and recall for sanitisation.}
\label{tab:precrecall}
\center
	\begin{tabular}{|l|r|r|}
		\hline
			 program & precision & recall \\ \hline
biojava & 0.996 & 0.976 \\
jabref & 0.982 & 1 \\
jmeter & 1 & 1 \\
jsoup & 1 & 1 \\
pdfbox & 1 & 1 \\
swagger-parser & 1 & 1 \\  \hline
	\end{tabular}
\end{table}

We discuss selected results in more detail in the next section.

\subsection{Discussion}
\label{ssec:evaluation:discussion}

\subsubsection{\textit{biojava}}

In \textit{biojava}, there are 17 failures and 281 errors when the network is down. Sanitisation successfully picks up all these tests except for eight (one failure and seven errors). Closer inspection reveals that six of these tests are in the class \texttt{TestMultipleAlignmentWriter} (package names omitted for brevity), and the network-related error occurs in the constructor for the class. Fixture setup in the constructor of test classes is not a good practice, and a better fix for this would be to refactor the test class that would also allow sanitisation to intercept it. The other false negative, which is a test error is in \texttt{TestProteinSuperposition}, which is caused by a network error that the sanitisation fails to intercept (in a method annotated with \texttt{@BeforeClass}). The test failure, which is also a false negative occurs in \texttt{File\-Down\-load\-UtilsTest.URLMethods::pingGoogleOK}. This illustrates the second source of false negative discussed in Section~\ref{ssec:network-dependencies:implementation} -- an \texttt{UnknownHostException} is thrown in a native method (\texttt{Inet6AddressImpl::lookupAllHostAddr}),  and the test swallows the exception (stacktrace) (\texttt{File\-Down\-load\-Utils::ping}
returns \texttt{false} when an exception is encountered).  As a result, the exception cannot be sanitised. There is also a false positive, \texttt{PDBStatusTest::testGetCurrent}, which is the test that fails in the baseline configuration. This test accesses the network but the failure is due to unexpected data from a REST API. %With these results, precision is 0.996 and recall is 0.98.

% @Shawn: old ICST results:
%9 FP , 1 FN
%FN detected: org.biojava.nbio.core.util.FileDownloadUtilsTest$URLMethods::pingGoogleOK
%FP detected: org.biojava.nbio.structure.align.util.AtomCacheTest::testGetStructureForChainlessDomains
%FP detected: org.biojava.nbio.core.search.io.blast.BlastXMLParserTest::testCreateObjects
%FP detected: org.biojava.nbio.structure.test.ecod.EcodInstallationTest::testDownloads
%FP detected: org.biojava.nbio.protmod.phosphosite.TestAcetylation::testAcetylation
%FP detected: org.biojava.nbio.structure.align.util.AtomCacheTest::testGetStructureForDomain1
%FP detected: org.biojava.nbio.structure.align.util.AtomCacheTest::testGetStructureForDomain2
%FP detected: org.biojava.nbio.structure.align.util.AtomCacheTest::testGetStructureForDomain3
%FP detected: org.biojava.nbio.structure.TestAtomCache::testGetScopDomain
%FP detected: org.biojava.nbio.structure.test.ecod.EcodInstallationTest::testGetStructure

%However, of those 67 do not succeed for at least one run when the network is on and sanitation is not used. I.e. the notions of precision and recall we have used are based on the assumption that we do have a clear run where in a baseline configuration (network on, no sanitation) all tests succeed.

\subsubsection{\textit{jabref}}

%0 FNs detected
% 1 FPs detected
%	FP detected: org.jabref.logic.remote.RemoteSetupTest::testPortAlreadyInUse()

For \textit{jabref}, there are 56 relevant tests and also 1 failure in the baseline configuration. The failure in the baseline configuration is also sanitised, which we observe as the single false positive. The particular false positive is a single test, \texttt{RemoteSetupTest::testPortAlreadyInUse}, which fails and when \textit{saflate} sanitisation is present, it is skipped. The failure is not network related and it is due to a bug in the test's use of \textit{mockito}. As all relevant tests are skipped by the sanitisation, there are no false negatives, leading to perfect recall and a precision of 0.982 due to the false positive.

%Reliance on network resources does not only mean that the network is enabled, but also that the respective servers are available.
%For example, even when the network is available,  36 tests fail in the database category alone (\texttt{gradle :databaseTest}) with a \texttt{org.postgresql.util.PSQLException} caused by a network exception indicating that the network server is unavailable. This is an issue with the fixture that requires some additional manual set up~\footnote{\url{https://devdocs.jabref.org/getting-into-the-code/testing}}.  This is reflected in a low precision as the metric is based on baseline assumption that all tests succeed when the network is available, which is clearly not the case here.
%, resulting in 73 false positives with respect to our precision metric.

%We also observe 4 false negatives, leading to a recall value of less than 1.  Debugging reveals that network exceptions are thrown and recorded but not in the thread used to execute the tests. For instance,  in \texttt{FulltextFetchersTest::higherTrustLevelWins} network access happens asynchronously using the \texttt{JabRefExecutorService} utility.  The false negatives occur as we have enabled \textit{saflate} concurrency support for \textit{jabref}, as there are a some test classes in \textit{jabref} that use  concurrent tests.  This has been discussed in Section~\ref{ssec:network-dependencies:implementation}.

\subsubsection{\textit{jmeter}}

There are 3 relevant tests in \textit{jmeter}, all of which are failures in the Spock specification, \texttt{DNSCacheManagerSpec}. An additional failure is also present in the baseline configuration. Sanitisation successfully skips all 3 relevant cases. There are no false positives, resulting in perfect recall and precision for the program with sanitisation.

\subsubsection{\textit{jsoup}}
In \textit{jsoup} there are 3 skipped tests in the baseline and one test case, \texttt{ConnectTest::handlesUnknown\-EscapesAcrossBuffer}, that terminates with an error when the network is off. This test is skipped with sanitisation on.

\subsubsection{\textit{pdfbox}}

There are 35 relevant tests (i.e., tests that should be sanitised) as these tests result in error when the network is off whilst they succeed with the network on. The sanitisation successfully detects these tests and skips them. The two additional skipped tests, which are already in the baseline run (network on, sanitisation off) are tests in \texttt{PDFBoxNonHeadless\-Test}  guarded by the assumption \texttt{assume\-False(Gra\-phics\-En\-viron\-ment.\-is\-Headless())} that fails when running the tests in a docker container. This is a good use of assumptions.

\subsubsection{\textit{swagger-parser}}

For \textit{swagger-parser}, the two cases which fail when the network is off are successfully skipped with \textit{saflate} on, there are no false positives resulting in perfect precision and recall.

\subsection{Performance}

The overhead of using \textit{saflate} is generally low. Performance data (in terms of runtime) is included in tables \ref{tab:unsanitized} and \ref{tab:sanitized}. The values reported are averages across runs. There are significant differences in the overhead that \textit{saflate} adds and the overhead added depending on whether the network is on or off. This depends on multiple factors such as the speed of network access when the network is available, timeout settings for connections and reconnect set up when the network is unavailable. We observed average overheads of 14\% when the network is on and 25\% when the network is off.  For three programs, \textit{jabref}, \textit{jmeter} and \textit{pdfbox}, the overhead is less than 10\% for both configurations.

We deem this as acceptable, and further optimisations are possible, such as managing timeouts and reconnect settings in the application, or using targetted annotations instead of the global deployment of \textit{saflate} for \textit{JUnit5}.

\section{Related Work}
\label{sec:relatedwork}

% \begin{itemize}
%   \item Mark Harman's positioning paper on the problems in identifying flaky tests \cite{harman2018start}
%   \item strategies to deal with flaky tests at google \cite{googleFlaky2016}
%   \item DeFlaker \cite{bell2018deflaker}
%   \item A container-based infrastructure for fuzzy-driven root causing of flaky tests \cite{Terragni2020container}
%   \item Understanding Reproducibility and Characteristics of Flaky Tests Through Test Reruns in Java Projects \cite{Lam2020Understanding}
%   \item FlakyLoc: flakiness localization for reliable test suites in web applications \cite{Mor2020FlakyLoc}
%   \item Concurrency-related flakiness in Android apps \cite{Dong2021Concurrency,Silva2020Shake}. Also an empirical study on flaky tests in android apps.
%   \item Detecting Flaky Tests in Probabilistic and Machine Learning Applications \cite{Dutta2020Detecting}
%   \item flaky coverage \cite{Vaidhyam2020Quantifying}
%   \item fixing order-dependent flaky tests with iFixFlakies \cite{Shi2019iFixFlakies}
%   \item Longitudinal Study of Flaky Tests \cite{Lam2020Longitudinal}.
% \end{itemize}

\subsection{Flaky Test Detection Techniques}
\label{ssec:relatedwork:techniques}

There are a number of techniques that have been proposed to detect flaky tests (mostly by looking at the change of the test outcome), with some of these tools able to identify the cause of flakiness. Often, tools focus on specific causes of flakiness, such as concurrency or order-dependency.

DeFlaker \cite{bell2018deflaker} attempts to avoid expensive reruns by detecting if test failures are due to flaky test behaviour without rerunning the entire test suite. DeFlaker works by recording the coverage of latest code changes and flags any newly failing test that did not exercise changed code as flaky.
FlakyLoc \cite{Mor2020FlakyLoc} does not detect flaky tests per se, but identifies causes for a given flaky test. The tool executes a flaky test in different environment configurations. These configurations are composed of  environment factors (e.g., memory sizes, CPU cores, browsers and screen resolutions) that are varied in each execution. The results are analysed using a spectrum-based localisation technique \cite{wong2016survey} that assigns a suspiciousness value and ranking to each configuration to determine the most likely factors that caused the flakiness.
RootFinder \cite{Lam2019Root} identifies potential causes of flakiness as well as the location in code that caused the flakiness. The tool can identify flakiness due to nine different causes, including network, time, IO, randomness, floating point operations and test order dependency. To do this, the tool instruments API calls during test execution, which can log interesting values (time, context, return value) and also add additional behaviour (e.g., add delays to trigger concurrency-related issues). %Post-execution, the logs are analysed by evaluating predicates (e.g., if the return value was the same at this point compared to previous times) at each point where it was logged. Predicates that evaluate to consistent values in passing and failing runs are likely to be useful in identifying the root causes.

Several tools target concurrency as a source of flakiness. FlakeShovel \cite{Dong2021Concurrency} targets event races that can cause test flakiness by  exploring different yet feasible event execution orders, however this is limited to GUI tests in Android apps.
\textit{Shaker} \cite{Silva2020Shake} exposes flakiness by adding noise to the environment in the form of tasks that also stress the CPU and memory whilst the test suite is executed.
NodeRacer \cite{endo2020noderacer} is a tool for JavaScript programs running on  \texttt{node.js} with  which accelerates manifestation of event races that can cause test flakiness. It uses instrumentation and builds a model consisting of a happens-after relation for callbacks.

Another group of tools is designed to detect flakiness due to order-dependent tests. iDFlakies \cite{lam2019idflakies}  uses reruns with randomised execution order, classifies all found flaky tests into either order-dependent or non-order dependent  tests. %The study based on iDFlakies is also an empirical study of test flakiness, as previous studies focussed on mining fix commits for flakiness or flakiness due to implementation-dependent (ID) tests (assumptions about underdetermined APIs).
DTDetector \cite{zhang2014empirically} presents four algorithms to identify test dependencies, which is manifested in test outcomes: reversal of test execution order, random test execution order, exhaustive bounded algorithm (which executes bounded subsequences of the test suite instead of trying out all permutations) and finally, the dependence-aware bounded algorithm that only tests subsequences that have data dependencies. %ElectricTest \cite{bell2015efficient} checks for data dependencies between tests by using a more sophisticated checks. Whereas DTDetector only checks for writes/reads to static fields, ElectricTest checks for changes to any memory reachable from static fields. PRADET  \cite{gambi2018practical} uses a similar technique to check for data dependences, but it also refines the output by checking for manifest dependencies, i.e. data dependence that also influences flakiness in test outcomes.
iFixFlakies \cite{Shi2019iFixFlakies} aims to automatically fix order-dependent tests by utilising helper tests to reset the state required for order-dependent tests to pass.
NonDex \cite{gyori2016nondex} targets implementation dependencies caused by assumptions developers make about under-determined APIs in the Java standard libraries, for instance the iteration order of hashed collections.

Herzig et al. propose using association rules to identify false test alarms \cite{herzig2015empi} where test cases are composed of test steps and failing patterns of test steps leading to test failure imply a false test alarm. This is similar to our approach, with the difference that we directly enforce it with a runtime component.

While we do not directly target source of flakiness like concurrency, test order dependencies and under-determined APIs, those issues often come to the fore in certain environments, and therefore assumptions checking for such environments can be used to disable tests and therefore deal with flakiness as suggested in Section~\ref{sec:approach}.
%But more importantly, \textit{saflate} proposes a framework to action the insights gained from some of these tools, and is in this respect orthogonal to those efforts. For instance, order dependencies could be actioned by a sanitiser based on a model of known test dependencies, and sanitise tests based on which other tests have already been executed, or not.

\subsection{Empirical Studies on Test Flakiness}
\label{ssec:relatedwork:empirical}

There have been several  empirical studies that investigated test flakiness and their root causes in open-source projects. The earlier study of Luo et al. \cite{luo2014empirical} investigated the common causes and fixing strategies of flaky tests by mining test fixes from Apache projects' commit history. The study showed that concurrency, order-dependency, resource leaks and network dependency are responsible for most of the flaky tests observed.\\
A similar work by Gruber et al., \cite{gruber2021empirical} studied the presence of flaky tests in Python projects and found that order-dependency is, by far, the most common cause - responsible for ~59\% of the flaky tests in these projects. Other non-order-dependent tests are predominantly caused by network and randomness APIs. The study also noted that a typical rerun to detect flakiness is not an effective method. %A test might need to be executed at least 170 times in order to be sure (with 95\% confidence level) that it is not flaky due to non-order-dependent reasons.\\
A recent study \cite{Hashemi2022flakyJS} investigated the major causes and fixes of flaky tests in JavaScript projects, noting that concurrency is the major know causes of flakiness in JavaScript. It was also found that flakiness due to order-dependency is not as prevalent in JavaScript as in Python \cite{gruber2021empirical} or Java \cite{luo2014empirical}. Thorve et al., \cite{Thorve2018empirical} identified five major causes of flakiness in Android application: concurrency, external dependencies, program logic, network, and UI, with concurrency bugs (36\% of the flaky tests commits) being the major cause of flaky tests.

\subsection{Flakiness caused by Network Dependencies}
\label{ssec:relatedwork:network}

%\todo[inline]{TODO-Amjed}
Network dependency (including connections, availability, and bandwidth) has been acknowledged as one of the key source of test flakiness \cite{luo2014empirical,eck2019understanding,parry2021survey}.
This includes both local and remote network issues. Local issues pertain to managing resources such as sockets (e.g. contention with other programs for ports that are hard-coded in tests), while remote issues concern failures in connecting to remote resources (e.g., attempting to reach a web server).
In a study of Python projects, Gruber et al. \cite{gruber2021empirical} noted that 413 tests are flaky due to network issues. It is also reported that 14\% of the flaky tests found in six large-scale Microsoft projects are due to network related causes \cite{lam2020study}. In a study consisting of Android projects \cite{Thorve2018empirical}, network is identified as a cause of flakiness of 8\% of the detected flaky tests.

Flakiness caused by network instability or availability has been pointed at as a major issue for developers in two separate developers’ surveys \cite{eck2019understanding,ahmad2019empirical}. Ahmad et al. \cite{ahmad2019empirical} noted that one common strategy to reduce potential flakiness in tests that utilise network resources is by intentionally undermining network infrastructure for ``worst case scenarios'' (e.g., assumes that a connection to an external resource cannot be established or a port is not available). This way the test is prepared to handle potential flakiness caused by network issues by design.
%Mor{\'a}n et al. \citeS{S99} studied network bandwidth in localizing flakiness causes.

\subsection{Assumption Inference and Flakiness Control in the Wild}
\label{ssec:relatedwork:inthewild}

\textit{XUnit.SkippableFacts} for xUnit.net~\footnote{\url{https://github.com/AArnott/Xunit.SkippableFact}} is similar to our network dependency extension for .NET written in C\# in that it can skip tests if certain exceptions occur.  It will not handle \textit{failed} tests though.

\textit{Unruly}~\footnote{\url{https://github.com/unruly/junit-rules}} is a set of JUnit4 rules to deal with flaky tests. \texttt{QuarantineRule} simply retries failing tests until they pass or a maximum number of failure is reached. This only deals with some very specific causes of flakiness that occur in the same environment (randomness, concurrency). \texttt{ReliabilityRule} is the opposite -- test only pass if the same test consistently  passes a number of times.

The conditional test annotations in \textit{JUnit5} in \texttt{org.junit.jupiter.api.condition} are closely related to our approach.  The difference is that we aim to infer the conditions during speculative executions on-the-fly, whereas those assumptions can be checked before the test in being executed.

The closest existing technique we are aware of are the \texttt{@PendingFeature} and the \texttt{@NotYetImplemented} annotations in \textit{spock} and \textit{groovy} respectively, already briefly discussed in Section~\ref{ssec:approach:staticvsdynamic}. The semantics of those annotations is based on speculative executions where test failures are sanitised as they are seen as tolerable during development. Interestingly, for \texttt{@PendingFeature}, the opposite is also true -- passes are converted into failures to indicate that the annotation has become redundant. This is closely aligned with the philosophy of test-driven development~\cite{beck2003test} where passing tests are used in the \textit{definition-of-done}.

%We do envisage a scenario where \textit{saflate}, and in particular the provenance information associated with the classifier we compute, can be used by engineers to add those standard annotations.  This step could also be automated by mechanically adding the respective standard annotations to tests.

\section{Conclusion}

We have presented a novel approach to address test flakiness by inferring assumptions which can sanitise tests that fail only due to environmental conditions that cannot be controlled in tests.

% \todo[]{I'd not call those not meaningful tests. they are still useful in a way as they show an issue with the test (e.g., test is dependent on network availability)} in the context of a  test execution. We have demonstrated that this can be done on-the-fly within the current build.
% Jens: I have rephrased this
%We have underpinned this approach conceptually with the notion of strong flakiness, arguing that this is the kind of flakiness most relevant to modern software engineering practices relying on automated builds.

We have illustrated the utility of this approach and our proof-of-concept implementation \textit{saflate} on six popular real-world Java and Groovy programs with tests relying on network availability. %, and with known issues related to this.
The results suggest that our method can achieve sanitisation with both high precision and recall, while the overhead to developers for using our method is considerably low, as it only requires adding a dependency and annotations to tests depending on networked resources. If \textit{JUnit5} is used, even the use of annotations can be avoided and \textit{saflate} can be deployed with only minor changes to the build scripts configuring the tests. Our results also suggest that the computational overhead imposed by our method is also low.

%This can then be used to avoid the unsatisfactory choices which developers currently have.

There are alternative use cases for our general approach such as existing sanitisation techniques for pending and not yet implemented features, therefore we believe that this method has some potential usage beyond dealing with network issues. A common theme between the use cases we have identified is instability that cannot be controlled and checked by engineers, in this respect, dealing with network dependencies and features that have not been completely implemented is similar.
%Sanitisation of tests depending on other tests is another possible application for the method proposed here.
%\todo[]{this last sentence is not very clear to me}
%Jens: I have removed this now

We think that the technique we have proposed here can be  useful in practice for projects that have \textit{some} tests with network dependencies. If a large number of tests are being sanitised (e.g., in an application with heavy use of network dependencies), our method might not be appropriate for such a project.  But for the programs we have investigated, the number of affected tests is relatively small, and often relate to marginal features. Then it can be argued that those tests failing should not block builds. At least, sanitised tests can be flagged and handled differently from failing tests in subsequent processes. For this reason, we believe that our method can add real value to industry practice.

\section*{Acknowledgment}
This work was funded by a New Zealand SfTI National Science Challenge grant No. MAUX2004. The work of the first author was also supported by Oracle Labs, Australia.

\Urlmuskip=0mu plus 1mu
\bibliographystyle{IEEEtran}
\bibliography{bibliography}

\end{document}